\documentclass[aps,prl,twocolumn,superscriptaddress]{revtex4-1}

\usepackage{amsmath}
\usepackage{amsfonts}
\usepackage{amssymb}
\usepackage{dcolumn}
\usepackage{gensymb}
\usepackage[utf8]{inputenc}
\usepackage[english]{babel}
\usepackage{epsfig}
\usepackage{graphics}
\usepackage{xcolor}

\usepackage{newtxtext} 
\usepackage{newtxmath}

\usepackage{microtype}
\usepackage{booktabs}
\usepackage{siunitx}

\begin{document}

\title{Diffractive Guiding of Waves by a Periodic Array of Slits}

\author{Dror Weisman}
\affiliation{School of Electrical Engineering, Iby and Aladar Fleischman Faculty of Engineering, 
Tel Aviv University, Tel Aviv 69978, Israel}

\author{C. Moritz Carmesin}
\affiliation{Institut f\"ur Quantenphysik and Center for
Integrated Quantum Science and Technology ($\it IQ^{ST}$), Universit\"at Ulm, Albert-Einstein-Allee 11, 89081 Ulm, Germany}

\author{Georgi Gary Rozenman}
\affiliation{School of Electrical Engineering, Iby and Aladar Fleischman Faculty of Engineering, 
Tel Aviv University, Tel Aviv 69978, Israel}
\affiliation{Raymond and Beverly Sackler School of Physics $\&$ Astronomy, Faculty of Exact 
Sciences, Tel Aviv University, Tel Aviv 69978, Israel}

\author{Maxim~A.~Efremov}
\affiliation{Institut f\"ur Quantenphysik and Center for
Integrated Quantum Science and Technology ($\it IQ^{ST}$), Universit\"at Ulm, Albert-Einstein-Allee 11, 89081 Ulm, Germany}
\affiliation{Institute of Quantum Technologies, German Aerospace Center (DLR), S\"oflinger Straße 100, 89077 Ulm, Germany}

\author{Lev Shemer}
\affiliation{School of Mechanical Engineering, Faculty of Engineering, Tel Aviv University, Tel Aviv 69978, Israel}

\author{Wolfgang P. Schleich}
\affiliation{Institut f\"ur Quantenphysik and Center for
Integrated Quantum Science and Technology ($\it IQ^{ST}$), Universit\"at Ulm, Albert-Einstein-Allee 11, 89081 Ulm, Germany}
\affiliation{Institute of Quantum Technologies, German Aerospace Center (DLR), S\"oflinger Straße 100, 89077 Ulm, Germany}
\affiliation{Hagler Institute for Advanced Study at Texas A$\&$M University, 
Texas A$\&$M AgriLife Research, Institute for Quantum Science and Engineering (IQSE), and Department of Physics and Astronomy, 
Texas A$\&$M University, College Station, TX 77843-4242, USA}

\author{Ady Arie}
\affiliation{School of Electrical Engineering, Iby and Aladar Fleischman Faculty of Engineering, 
Tel Aviv University, Tel Aviv 69978, Israel}


\begin{abstract}
We show that in order to guide waves, it is sufficient to periodically truncate their edges. The modes supported by this type of wave guide propagate freely between the slits, and the propagation pattern repeats itself. We experimentally demonstrate this general wave phenomenon for two types of waves: (i) plasmonic waves propagating on a metal-air interface that are periodically blocked by nanometric metallic walls, and (ii) surface gravity water waves whose evolution is recorded, the packet is truncated, and generated again to show repeated patterns. This guiding concept is applicable for a wide variety of waves. 
\end{abstract}

\maketitle

Guiding waves is a challenge scientists and engineers have grappled with for more than a century \cite{Rayleigh1897, Packard1984}. Although waveguides come in a variety of geometries and are made from numerous materials \cite{Okamoto2006} they share one common feature: The underlying guiding mechanism is based on a spatial confinement of the wave in the transverse direction. In this Letter we study a new guiding concept where rather than confining the wave, guiding is achieved by a periodic truncation of its edges using a periodic array of slits. \\
The motivation for our work originates from the phenomenon of diffractive focusing \cite{Weisman2017,Case2012,Goncalves2017,Vitrant2012}, in which a plane wave that passes through a single slit is first focused before it expands. Here, the intensity increases up to a factor of 1.8 (4) in the case of a vertical (circular) slit. A periodic array of these slits could therefore represent the \textit{diffractive} lens waveguide, in complete analogy to the well-known \textit{refractive} lens waveguides \cite{Marcuse1964}.\\
After every slit, the wave propagates freely, until it reaches the next slit where the edges are chopped off. This elementary mechanism \cite{CommentStutzle} supports different eigenmodes, and the lowest mode can be guided over significant distances with low propagation losses. Moreover, the on-axis intensity can be significantly higher in comparison to the freely propagating non-truncated case, even though each slit removes part of the overall energy of the entire wave. Diffractive guiding is therefore an attractive solution whenever it is difficult to realize traditional waveguides, for example, for electromagnetic waves in the terahertz regime \cite{Geloni2011}.\\
In our Letter we study theoretically and experimentally the phenomenon of diffractive guiding of surface plasmon polaritons (SPPs) in \textit{space}, as well as surface gravity water waves (SGWW) in \textit{time}. For SPPs we succeeded in measuring the repeated propagation pattern of the first and second eigenmodes created by a periodic array of slits made from nanometric metallic walls. For SGWW we obtain the propagation through several slits by recording the full wave packet data, truncating the edges of the wave data and starting a new measurement with the truncated wave. We emphasize that many other types of waves, including electromagnetic waves, sound waves or matter waves can be guided using the same method.\\
The time evolution of paraxial plasmonic beams, like other electromagnetic waves, follows from the paraxial Helmholtz equation \cite{Epstein2016}, which is formally equivalent to the free Schrödinger equation. Similarly, the propagation of surface gravity water waves under appropriate conditions is governed by an equivalent equation \cite{Rozenman2019a}.\\
In order to study guiding, we assume perfectly absorbing slits, no reflections and no losses. We decompose the array of slits into units consisting of one slit and free propagation till the next slit. 
The wave function $\psi=\psi(\chi, \tau)$ can thus be obtained from the incident wave $\psi_0(\chi_0)\equiv\psi(\chi_0, \tau=0)$ directly before the slit by the integral
\begin{equation}
    \psi(\chi, \tau)=\int_{-\infty}^\infty \mathrm{d} \chi_0 K(\chi, \tau|\chi_0)\psi_0(\chi_0),\label{propagantion_unit}
\end{equation}
where the propagator
\begin{equation}
 \begin{split}
 \label{Propagator}
 K(\chi,\tau|\chi_0)\equiv\frac{1}{\sqrt{i\tau}}\exp\left[i\pi\frac{(\chi-\chi_0)^2}{\tau}\right]\Theta\left(\frac{1}{2}-|\chi_0|\right)
 \end{split}
\end{equation}
is the product of the propagator of the free motion \cite{Case2012,Weisman2017,feynman2010} and the Heaviside step function $\Theta$ modeling the truncation of the wave function at the slit. Here, we use the dimensionless propagation coordinate $\tau$ as well as the transverse coordinate $\chi\equiv x/a$, which is scaled by the width $a$ of the slit.\\
For optical as well as plasmonic waves, $\tau$ translates into the propagation distance $z$ according to $\tau\equiv z\lambda/a^2$, where $\lambda$ is the wavelength \cite{Comment_propagatorSPP,Comment1}.
In the case of a matter wave \cite{feynman2010} we have $\tau\equiv ht/(ma^2)$, where $t$ denotes the time,  $m$ the mass of the particle and $h$ is Planck's constant.\\
The wave function after the array of slits can be computed by iterating Eq.~\eqref{propagantion_unit}, where the result of one integration serves as the source for the next one.

Since the wave function is cut at the slits, there is some intrinsic loss in diffractive guiding that depends on the spatial distribution of the wave function at the slit. This fact favors an analysis in terms of the eigenmodes of the system, since only modes with the lowest losses survive after passing through the array of slits.
The periodicity of the slits further reduces the problem to finding the eigenmodes of just one unit of the system.

Since the eigenmode repeats itself after propagation through one unit of the array, its wave functions from one slit to the next may differ only by a constant factor.
In order to obtain the eigenmodes and eigenvalues we discretize the $\chi$-coordinate and cast the integral in Eq.~\eqref{propagantion_unit} into a matrix vector multiplication.\\
Hence, we arrive at the eigenvalue equation
\begin{equation}
 \label{Eigenmodes_equation}
 \mathsf{K}_\tau \boldsymbol{\psi}_i = \mu_i \boldsymbol{\psi}_i,
\end{equation}
 where $\mathsf K_\tau$ is the propagator matrix for a separation $\tau$  between the slits, while $\boldsymbol{\psi}_i$ are the eigenmodes discretized in $\chi$. The corresponding eigenvalues $\mu_i$ are related to the transmission coefficients $T_i$ by $T_i=|\mu_i|^2$. The eigenmodes and eigenvalues are computed \cite{Comment2} by the numerical diagonalization of $\mathsf K$.

\begin{figure}
  \includegraphics[width=0.46\textwidth]{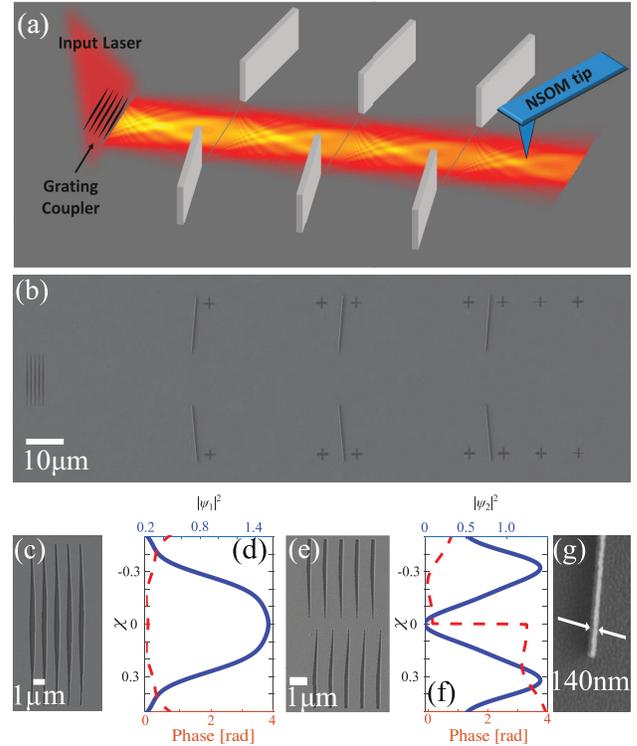}
  \caption{\label{Figure1} 
  Experimental setup for diffractive guiding of surface plasmon polaritons (SPPs): (a) Our device consists of a grating coupler for the excitation of the  mode, plasmonic walls for truncation of the wave, and a near-field scanning optical microscope (NSOM) tip. (b) Scanning electron microscope (SEM) image of our diffractive waveguide device with three plasmonic slits. (c) and (e) SEM images of the grating couplers used for the plasmonic excitation of the first and second modes, shown in (d) and (f) by the truncated initial amplitude and phase. (g) SEM image of a single metallic wall.}
  
\end{figure}

In order to test experimentally the guiding capabilities of  this device, we consider first the case of SPPs, which are electromagnetic surface waves propagating along a metal-dielectric interface coupled to collective charge oscillations inside the metal \cite{Maier2007}. Figure \ref{Figure1}(a) illustrates the corresponding experimental setup.\\
In order to periodically truncate the edges of the beam, we have used a plasmonic slit that consists of two silver metallic walls having a height of several hundred nanometers. Since these beams are strongly confined to the interface, most of the energy is either scattered or reflected at an angle with respect to the optical axis  from the tilted metallic wall, and as a result, the wave cannot pass. We have fabricated \cite{Comment_SPP_EXP} several structures having  a fixed distance between adjacent slits, but with a different number of slits on each structure. 

Figure \ref{Figure1}(b) displays such a structure with three successive plasmonic slits. In order to excite a specific mode supported by such an array, with a desired wavefront, we used a plasmonic near-field hologram coupler for controlling both the amplitude and phase of the excited plasmonic beam \cite{Epstein2014,Lee1979}. Scanning electron microscope (SEM) images of the gratings for the excitation of the first and second modes can be seen  in Figs.~\ref{Figure1}(c) and (e). The amplitudes (solid curve) and phases (dashed curve) of the corresponding  modes are displayed in Figs.~\ref{Figure1}(d) and (f).
In comparison to other plasmonic waveguides, our waveguide exceeds in terms of propagation length that of hybrid plasmonic waveguides and is comparable to dielectric loaded waveguide, although it is still less than long-range SPP waveguide \cite{Krasavin2015}. A quantitative comparison can be found in Table \ref{table:1}.

\begin{figure}[tb]
  \includegraphics[width=0.48\textwidth]{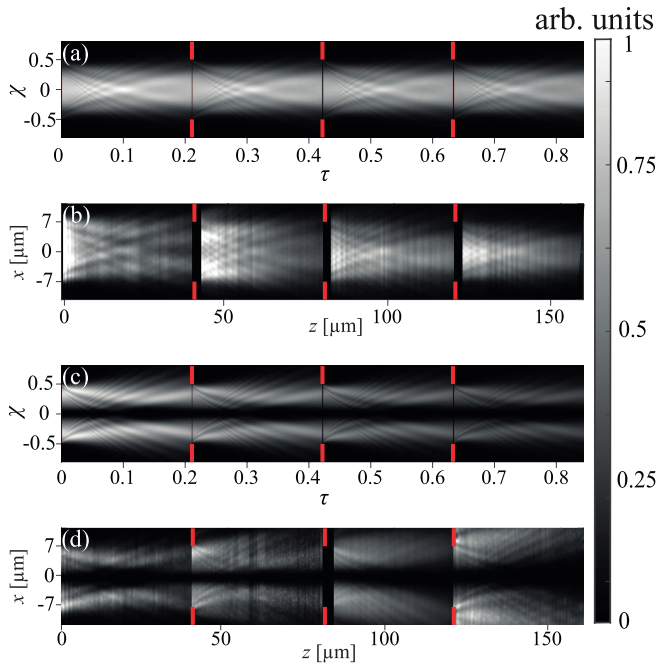}\\
  \caption{\label{SPPmodes} 
  Diffractive guiding in space: Theory versus experiments based on surface plasmon polaritons. Simulations of the propagation intensity patterns $|\psi|^2$ for the first (a) and second (c) mode. The vertical red lines represent the apertures truncating the wave.
  Experimental measurement of the plasmonic intensity distribution beyond the grating coupler and after passing through one, two and three slits, for the first (b) and second (d) modes. For a measurement of the first mode after passing through five slits see Supplemental Material. Each experimental plot is a combination of four different measurements taken with four different structures. The experimental intensity plots are in arbitrary units.}
    
\end{figure}

We have solved Eq.~\eqref{Eigenmodes_equation} for the separation $\tau=0.211$ between successive slits, corresponding to the experimental parameters $L=39.2\,\text{\textmu m}$ (the distance between successive slits),  $a=14\,\text{\textmu m}$ and $\lambda=1055\,\text{nm}$ neglecting losses, and present the resulting intensity distribution patterns of the propagating first and second mode in Figs.~\ref{SPPmodes}(a) and (c), respectively. 
The corresponding eigenvalues are $\mu_1=0.9621+0.1163i$ and $\mu_2=0.778 - 0.396i$ lead to the intensity transmission coefficients $T_1=0.94$ and $T_2=0.76$. It is noteworthy that the overall transmission of the first mode is quite high, despite the fact that part of the energy is truncated by the slits. Moreover, for the first mode we get a $20\%$ increase in the intensity at the focal point relative to the peak value in the intensity profile of the inserted eigenmode, due to diffractive focusing \cite{Weisman2017,Case2012,Goncalves2017,Vitrant2012}. \\
Notice that the simulations presented here do not take into account back propagation. Depending on the implementation, interference can appear during the propagation between the slits, due to back reflections and scattering from the edges.
We emphasize that because of plasmonic losses, the intensity transmission coefficients of the same eigenmodes are reduced from 0.94 to 0.86 and from 0.76 to 0.70, for the first and second eigenmodes, respectively, calculated for silver-air interface with the relevant silver permittivity \cite{Johnson} for the above wavelength. \\
Figure \ref{SPPmodes}(b) presents experimental measurements of the intensity distribution patterns for the \textit{first} mode propagating through a periodic structure of plasmonic slits, after it is excited by the modulated grating coupler shown in Fig. \ref{Figure1}(c), and after passing through one, two, three slits (four different structures) \cite{Supp}. In each case, we have measured the intensity distribution only after the last slit. In this way, we are not affected by reflections from the next slits, and we get a more precise measurement of the intensity distribution after each slit. The grating excites a pattern similar to the desired first mode, in excellent agreement with the simulation. Moreover, the pattern repeats itself after having passed through a new slit that truncates the sides of the wave.

Next, we show experimentally the excitation and guiding of the \textit{second} mode corresponding to the second largest eigenvalue, $\mu_2$ and supported by the same periodic structure of plasmonic slits. This mode has two main lobes, with a $\pi$-phase difference between them, as shown in Fig.~\ref{Figure1}(f). Similarly to the excitation of the first mode, we designed a specific grating coupler \cite{Epstein2014} for the second mode (see Fig. \ref{Figure1}(e)), and then measured the beam propagation through several slits as displayed in Fig.~\ref{SPPmodes} (d). Due to the phase difference between the two lobes, destructive interference takes place along the entire propagation axis at the center of the slit. It seems that for the second mode measurement after three slits, shown in the right most square in Fig. \ref{SPPmodes}(d), the plasmonic walls fail to completely block the SPP propagation and a relatively high intensity appears on the edges. 

Although we are not able to measure the transmission coefficient directly, since in each measurement we scan an area after a different number of slits and use different structures, we can still determine it by calculating which part of the intensity should remain, assuming we truncate the wave exactly in the position of the next slit and have an ideal slit. From the experimental measurements of the first mode we found the  transmission coefficients $0.9$, $0.89$, $0.91$, $0.95$, and $0.935$  by integrating  the intensity that lies within the slit relative to the total intensity, at the theoretical positions of slits number $1, 2, 3, 4$ and $6$, respectively. The average result is $0.917\pm0.026$, which is in good agreement with the theoretical value $T_1=0.94$. 

For the second mode, we obtain the values $0.88$, $0.77$, $0.61$, and $0.43$ of the transmission coefficient with respect to the positions of the slits 1, 2, 3 and 4,  which indicates that in the experiment, the confinement of the second mode is more difficult to achieve, as can be seen in Fig. \ref{SPPmodes}(d). The average value reads $0.67\pm 0.19$, while the theoretical value is $T_2=0.76$.

Diffractive guiding is a general wave phenomenon, and is not limited to plasmonic beams. We now present a different example for guiding which works in the \textit{time} domain and relies on surface gravity water waves. Indeed, the propagation of the SGWW envelopes also follows \cite{Rozenman2019, Dudley2014, Chabchoub2012, Chabchoub2013,Fu2015,Rozenman2019a} from the free Schrödinger equation. Hence, the  above eigenmode analysis applies, with the corresponding propagator for deep water waves  \cite{Comment_propagatorWater}.\\
For this purpose we modulate the envelope of the surface water gravity wave according to the first eigenmode. Therefore, the wave packet generated by the wave maker reads $\eta(t,x=0)=a_0A(t)\cos(\omega_0 t+\phi(t))$, with the carrier frequency reads $\omega_{0}= 10\,\text{rad}/\text{s}$ and the initial amplitude $a_{0}=4.9\,\text{mm}$. Here $A(t)$ and $\phi(t)$ represent the amplitude and phase of the eigenmode. \\
Then, the time-dependent elevation of the wave was measured using wave gauges at different positions in the tank and stored in a computer. This data was numerically  truncated  at the location of the next temporal slit, and was then sent to the wave maker for a new excitation based on the previous measurement. This way, we were able to cascade several slits in the time domain and observe the effect of diffractive guiding.

\begin{figure}[tb]
  \includegraphics[width=0.48\textwidth]{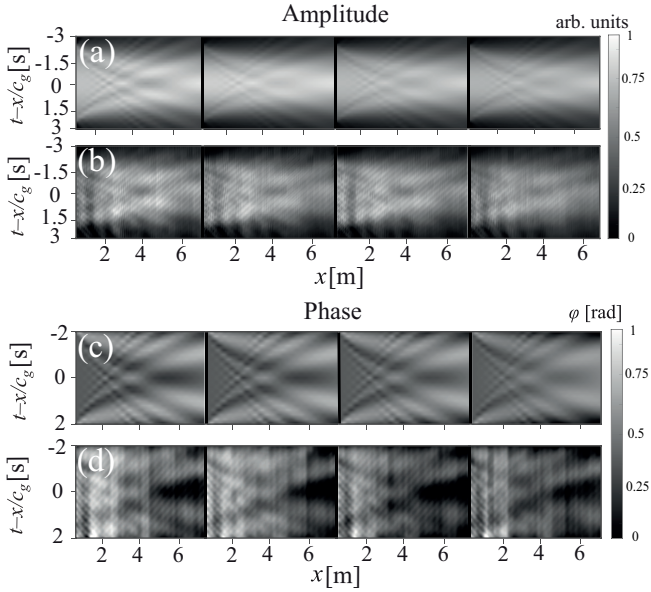}\\
  \caption{\label{Figure3} 
  Diffractive guiding in time: Theory versus experiment based on surface gravity water waves. Theoretical patterns (a)  and experimental measurements (b)  of the amplitude. Theoretical (c) and experimental (d) distributions of the phase, obtained from the measurements by the Hilbert transform \cite{Supp} in a frame of reference at moving with the velocity $c_{g}=0.49\,\text{m}/\text{s}$, and low steepness, $\varepsilon=0.05$. Each experimental plot is a combination of four different measurements.}
\end{figure}

Figure \ref{Figure3} shows our measurements for the water wave experiment, where the temporal slit has a width of $t_{0}=4.2\,\mathrm{s}$, with the period $L=8.5\,\mathrm{m}$, to match the parameters that were used for the simulation of Fig. \ref{SPPmodes}(a). We worked with low steepness waves ($\varepsilon=0.05$) so that the nonlinear terms of the wave equation \cite{Rozenman2020} can be neglected. \\
In our experiment we record the full temporal variation of the wave packet, thus enabling us to extract not only its amplitude, but also its phase. Figure \ref{Figure3} shows that the pattern repeats itself for both the intensity and phase measurements, as expected. The results are normalized in each run by the maximal intensity amplitude.

To emphasize the generality of the phenomenon for different types of waves, we present a quantitative comparison between losses in a diffractive guiding system and other known waveguides. Table \ref{table:1} shows several examples of three types of waves and compare between the propagation distance in our system relative to other proposed and explored waveguides.

\begin{table}[tb]

\centering

\sisetup{range-phrase = --, range-units=single}
\begin{tabular}{ cccccc} 
 \toprule
   & {$\lambda$/period} & {$a$} & {$L$} & {$L_\text{prop}$} & {Typical $L_\text{prop}$} \\ 
   \midrule
 SPPs & \SI{1.54}{\micro m} & \SI{20}{\micro m} &  \SI{60}{\micro m} & \SI{123}{\micro m} &  \SIrange{40}{90}{\micro m} \cite{Krasavin2015, Holmgaard2007}\\ 
 THz &  \SI{0.15}{mm} & \SI{2}{mm} &  \SI{6}{mm} & \SI{233}{cm} & \SIrange{1}{500}{cm} \cite{THz2013}\\ 
 SGWW &  \SI{1.59}{s} & \SI{4.2}{s} &  \SI{8.5}{m} & - & -\\ 
 
 \bottomrule
 
 \end{tabular}
 \caption{Comparison of diffractive guiding propagation distances ($L_\text{prop}$) and typical values of other waveguides for several types of waves. For SPPs, Gold-air interface for telecom wavelength $\lambda_0=\SI{1.55}{\micro m}$ was analyzed and dielectric loaded waveguides are presented for the values appeared in the typical values column. For THz waves, the calculation was made using 3D simulation, with two transverse axes, using a square aperture.}
\label{table:1}
\end{table}

\begin{figure}[tb]
  \includegraphics[width=0.48\textwidth]{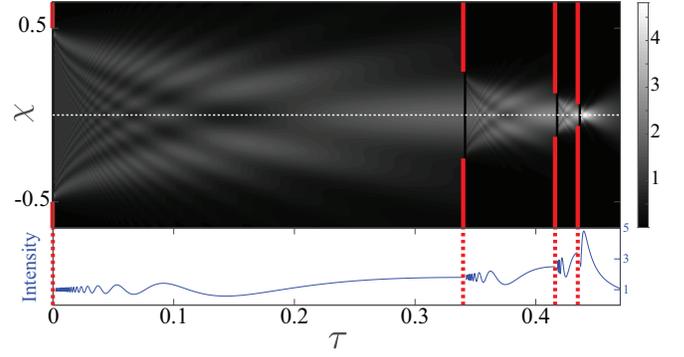}\\
  \caption{\label{Figure4} 
  Diffractive focusing and guiding for tapering. Intensity distribution simulation of the propagation of a plane wave through an array of slits, where each slit is located at the focal plane and has a width that equals to half of the former slit. The inset shows the intensity profile along $\chi=0$, the white dashed line. The red lines represent the positions of the slits.}

\end{figure}

The concept of cascading several slits can be also used with a different scheme in order to achieve higher intensity and a narrower spatial confinement. Figure \ref{Figure4} shows such a structure that consists of four slits, each slit has half the width of the previous slit. As a result, an initial plane wave is focused repeatedly. This type of structure can be applied for tapering purposes, to couple waves to a system with better efficiency. It can also be located at the end of the periodic array of slits to maximize the intensity at the end and to connect between different systems.

We also investigate two more types of periodic structures. A periodic array of slits in which each unit cell consists of two different slits with different widths and different propagation distances, and a periodic array of slits where successive slits are shifted. For each of these structures we succeeded to use the same method that we showed here in order to find the eigenmodes of the system. More details can be found in the Supplemental material.

In summary, we have shown a new type of wave guide based on the phenomenon of diffractive focusing, and have experimentally measured the effect in the \textit{spatial} domain using plasmonic beams, and in the \textit{time} domain employing surface gravity water waves. The propagation losses of our diffractive plasmonic waveguide are comparable to existing state-of-the-art guiding devices, e.g. dielectric loaded waveguides. Unlike other guiding systems, the eigenmodes evolve also in the transverse direction during the propagation. Nevertheless, we recover after propagation in and truncation after each unit cell, the same pattern. \\
This concept of guiding using a periodic structure of slits can be applied to many different types of waves such as matter waves, electromagnetic waves, sound waves, etc. Specifically, the method may be useful in several regions of the electromagnetic spectrum, such as THz regime \cite{Geloni2011}, where it is quite difficult to realize a waveguide by other means. Moreover, we emphasize that this approach of cascading slits can help to achieve a higher energy concentration at specific points compared to free propagation.\\
Finally we note that the group velocity in such a system is close to that of a freely propagating wave. We therefore believe that diffractive guiding can be useful in a wide range of applications.\\

\begin{acknowledgments}

We thank Danveer Singh and Dolev Roitman for their help with the fabrication of the grating couplers, Tamir Ilan for technical support and assistance and Prof. Avi Gover for helpful discussions. This work is supported by the German-Israeli DIP Project (AR 924/1-1, DU1086/2-1), the Israel Ministry of Science, Technology and Space (Grant No. 3-12473) and the ISF (Grant No.508/19). M.A.E. is thankful to the Center for Integrated Quantum Science and Technology ($\it IQ^{ST}$) for its generous financial support. W.P.S. is grateful to Texas A$\&$M University for support through a Faculty Fellowship at the Hagler Institute for Advanced Study at the Texas A$\&$M University as well as to the Texas A$\&$M AgriLife Research. The research of the $\it IQ^{ST}$ is financially supported by the Ministry of Science, Research and Arts Baden-Württemberg.

\end{acknowledgments}




\begin{thebibliography}{99}
\bibitem{Comment_propagatorSPP} Using the relations $\tau\equiv z\lambda/a^2$, $\chi\equiv x/a$, we obtain the propagator $K_\text{SPP}(x,z|x_0)\equiv\frac{1}{\sqrt{i\lambda z}}\exp\left[i\pi\frac{(x-x_0)^2}{\lambda z}\right]$ of the paraxial Helmholtz equation for plasmonic beams.
\bibitem{Comment_SPP_EXP} In our plasmonic experiments the size of the slit is $a=\SI{14}{\micro m}$, the distance between each slit is $L=\SI{39.2}{\micro m}$, and the SPP's wavelength is $\lambda_\text{SPP}=\SI{1.055}{\micro m}$ since it is excited by a $\SI{1.064}{\micro m}$ CW laser and is propagating along a silver-air interface. These parameters correspond to the simulations presented in Fig. \ref{Figure1}(a,c) for $\tau=0.211$. The plasmonic walls were fabricated by a standard E-beam lithography process using PMMA-A7 resist, followed by $\SI{400}{nm}$ evaporation of silver and lift-off to create $\SI{400}{nm}$ height, $\sim\SI{150}{nm}$ thin, metallic wall (see Fig. \ref{Figure1}(g)). In order to reduce the amount of inner reflections inside the device, the silver walls were placed with a $5\degree$ tilt to send out reflected energy. The plasmonic intensity distribution was measured by NSOM.
\end{thebibliography}

\newpage

\end{document}